\documentclass[aps,preprint,superscriptaddress,amsfont,graphicx,nofootinbib,preprintnumbers]{revtex4}%
\usepackage{color,graphicx,epsfig}
\usepackage{ifpdf}
\usepackage{amsmath}
\usepackage{bm}
\usepackage{color}
\usepackage[english]{babel}
\usepackage{graphicx}%
\usepackage{amsfonts}%
\usepackage{amssymb}
\usepackage{braket}
\usepackage{hyperref}
\usepackage{enumerate}

\definecolor{nicered}{rgb}{0.7,0.1,0.1}
\definecolor{nicegreen}{rgb}{0.1,0.5,0.1}
\hypersetup{colorlinks,citecolor= nicegreen,linkcolor= nicered}

\newcommand{\be}{\begin{equation}}
\newcommand{\ee}{\end{equation}}
\newcommand{\bea}{\begin{eqnarray}}
\newcommand{\eea}{\end{eqnarray}}

\definecolor{Red}{rgb}{1.,0.,0.}

\def\OMIT#1{}

\begin{document}

\def\JHU{Department of Physics and Astronomy, Johns Hopkins University, Baltimore, USA}
\def\KIT{Institute for Theoretical Particle Physics, KIT, Karlsruhe, Germany}
\def\CERN{CERN Theory Division, CH-1211, Geneva 23, Switzerland}

\preprint{CERN-PH-TH-2015-192}
\preprint{TTP15-030}

\title{Fiducial cross sections for Higgs boson production in association with a jet at next-to-next-to-leading order in QCD}

\author{Fabrizio Caola}            
\email[Electronic address: ]{fabrizio.caola@cern.ch}
\affiliation{\CERN}

\author{Kirill Melnikov}            
\email[Electronic address: ]{kirill.melnikov@kit.edu}
\affiliation{\KIT}

\author{Markus Schulze}            
\email[Electronic address: ]{markus.schulze@cern.ch}
\affiliation{\CERN}

\begin{abstract}

We extend the recent computation of Higgs boson production in association with a jet through 
next-to-next-to-leading order in perturbative QCD by including decays of the Higgs boson
to electroweak vector bosons. This allows us to compute fiducial cross sections and 
kinematic distributions including realistic selection criteria for the Higgs boson decay products.
As an illustration, we present
results for $pp \to H + j \to \gamma \gamma + j$ closely following the ATLAS 8 TeV analysis and
for $pp \to H+ j \to W^+W^- + j \to e^+ \mu^- \nu \bar \nu + j $ in a CMS-like 13 TeV setup.

\end{abstract}

\maketitle

\section{Introduction} 
Studies of the Higgs boson discovered by the ATLAS and CMS 
collaborations~\cite{:2012gk,:2012gu} will be at the focus 
of the experimental program during Run II of the LHC. The interpretation of future measurements 
of Higgs boson production and decay rates  in terms of Higgs boson couplings to matter and gauge 
fields  and Higgs boson quantum numbers will 
rely on the comparison of  measured event rates and kinematic distributions 
with   results of theoretical  
modelling of such processes   in the Standard Model. 
It is hoped that such comparisons will help to elucidate the nature of the Higgs particle and explore the 
mechanism of electroweak symmetry breaking in detail \cite{atlascoup,cmscoup}. 

Recently, the ATLAS collaboration made an important  step 
forward in presenting the results of analysis of the Higgs boson production and decay at the LHC. 
Indeed, in contrast to many other Run I LHC measurements, 
the ATLAS collaboration  measured  fiducial volume 
cross sections and a variety of kinematic distributions 
  in the process  $pp \to H+{\rm jets}$ \cite{Aad:2014lwa}.  
The 
Higgs boson was identified through its decay to photons, $H \to \gamma \gamma$.

Fiducial volume measurements allow for a direct comparison between data and theoretical predictions 
minimizing extrapolation uncertainties. 
Although many of the ATLAS fiducial volume measurements
are currently limited by statistical uncertainties,  
this will certainly change in the current run of the LHC. 
Therefore, the important issue for  the near future  is  the 
availability of highly accurate theoretical predictions that 
can be used to describe complicated fiducial 
volume measurements.

We will now summarize the most advanced fixed order computations 
related to Higgs boson production and decay  at the LHC.
The inclusive 
production cross section of the Higgs boson has recently been 
computed through next-to-next-to-next-to leading order in perturbative QCD~\cite{nnnlo}.
This computation  refers to the total 
cross  sections and can not be used for the direct comparison with fiducial volume
measurements without extrapolation. The computation 
of $H+j$ production at the LHC 
has recently been  
extended to next-to-next-to leading order (NNLO) in perturbative QCD, in a fully differential manner 
\cite{hjetnnlo,hjetnnlo1}\footnote{For earlier partial results, see~\cite{Boughezal:2013uia,Chen:2014gva}.}. 
Unfortunately, in Refs.~\cite{hjetnnlo,hjetnnlo1} 
decays of the Higgs boson were not considered; for this reason the comparison of the 
results of Refs.~\cite{hjetnnlo,hjetnnlo1} 
with the results of the fiducial measurements  is also not possible. 
When the Higgs boson is produced in association with two and three jets, the NLO QCD 
computations provide the 
state-of-the-art results~\cite{Campbell:2010cz,vanDeurzen:2013rv,Cullen:2013saa}.
 In those NLO QCD computations decays of the 
Higgs bosons are routinely taken into account.
We note that the NNLO QCD computations of $pp \to H+j$ \cite{hjetnnlo,hjetnnlo1}
combine the NNLO prediction for the {\it exclusive} 
$H+j$ cross section with 
the NLO QCD prediction 
for the   exclusive $H+2j$ cross section 
and the LO prediction for the  $H+3j$ cross section, making 
them particularly  suitable for  studying  Higgs boson
production in association with  different number  of jets in a consistent way. 

It is relatively  straightforward to extend the fully-differential $pp \to H+j$ computation 
reported in Ref.~\cite{hjetnnlo} 
to include decays of the Higgs boson into electroweak gauge bosons 
since the Higgs boson is a spin-zero particle 
and no spin correlations need to be considered.  
This is what we do in this 
paper for a variety of the Higgs boson decay modes. 
Once this is done, it becomes possible to calculate  
fiducial volume cross sections and kinematic distributions and
directly compare with experimental measurements. 
The very fact that it is possible to do that 
through next-to-next-to-leading order in the expansion in the strong 
coupling constant,  represents an impressive  milestone in an application 
of perturbative QCD to the description of hard collisions at the LHC. 

Our paper is organized as follows. 
In Section~\ref{sect2}, we briefly summarize the theoretical and experimental setup. 
In Section~\ref{sect3}, we present the results for 
fiducial volume cross sections and kinematic distributions at the $8$~TeV LHC for the $H \to \gamma \gamma$ 
decay mode and at  the $13$~TeV LHC for the $H \to WW^* \to e^+\mu^- \nu \bar \nu $ decay mode.   We also compare the 
 results of the fiducial volume computation 
for the $\gamma \gamma$ final state  with the results 
of the ATLAS measurement.  We conclude in  Section~\ref{conc}.

\section{The setup}
\label{sect2}

\subsection{Theory}

We begin by summarizing  the theoretical framework that we use  in the computation. 
We work in an effective field theory obtained by integrating out the top quark. 
We employ the method of 
improved sector decomposition developed  in Refs.~\cite{Czakon:2010td,Czakon:2014oma,Boughezal:2011jf}.
This method is  based on the factorization of scattering amplitudes  in soft and collinear limits 
and on particular way of splitting the phase-space into sectors, where  soft and collinear 
singularities are easily  identified. 
For this calculation, we require a large number of matrix elements that are used to construct differential cross sections. 
In particular,  we need 
the two-loop virtual corrections to the partonic channels $gg \to Hg$ and $qg \to Hq$; 
the one-loop virtual corrections to $gg \to Hgg$, $gg \to Hq\bar{q}$, $qg \to Hqg$, $q\bar{q} \to H Q\bar{Q}$ and 
 the double real emission processes 
$gg \to Hggg$, $gg \to Hgq\bar{q}$, $qg \to H qgg$ and $qg \to HqQ\bar{Q}$, where the $Q\bar{Q}$  pair can be of any flavor.  The helicity amplitudes for 
all of these processes are available in the literature. The 
two-loop amplitudes were computed  in  Ref.~\cite{Gehrmann:2011aa}. 
The one-loop corrections to the four-parton processes are known~\cite{gghgg_1loop}. For five-parton tree-level amplitudes, we use
compact results obtained using BCFW recursions~\cite{simon_tree}.   

It is non-trivial to combine   processes with different particle multiplicities as  
required for any NNLO QCD computation. Our method  for 
doing that is described in \cite{Boughezal:2013uia};  we do not repeat that discussion here. 
As already mentioned in the Introduction, the inclusion of Higgs boson decays is straightforward 
since the Higgs boson is a spin-zero particle. The only technical issue that arises is a significantly 
larger  phase space that needs to be considered and the ensuing difficulties with the Monte-Carlo 
integration. However, the numerical challenges that appear in fiducial volume computations turn out to be not 
prohibitive. In particular, we find that for the $H\to \gamma \gamma$ decay mode we need roughly the same
amount of statistics as for the stable Higgs case, while for $H\to 4l$ the amount of statistics should
be increased by a factor between 2 and 4. 

\subsection{$H \to \gamma \gamma$}

We continue by listing the selection criteria employed by the ATLAS collaboration \cite{Aad:2014lwa}.  
We are interested in the process $pp \to H+j$,  where the Higgs boson decays 
to two photons.  Final state jets are defined using 
the anti-$k_\perp$ algorithm \cite{Cacciari:2008gp} 
with $\Delta R =0.4$ and 
$p_{\perp,j} > 30~{\rm GeV}$.
Jets are required to have rapidities $y_j$ in an interval $ -4.4 < y_j < 4.4$.  
The two photons from the Higgs decay 
must  have the transverse momenta 
$p_{\perp, \gamma_1} > {\rm max} \left ( 25~{\rm GeV}, 0.35~m_{\gamma \gamma} \right)$ and 
$p_{\perp, \gamma_2} > {\rm max} \left ( 25~{\rm GeV}, 0.25~m_{\gamma \gamma} \right)$, respectively, where 
$m_{\gamma \gamma}$ is the invariant mass of the two photons. In our calculation, 
the Higgs boson decays are described 
in the narrow width approximation, so that 
we always have $m_{\gamma \gamma} = m_H = 125~{\rm GeV}$. 
Then, the  
above conditions imply $p_{\perp, \gamma_1} > 43.75~{\rm GeV}$ and $p_{\perp, \gamma_2} > 31.25~{\rm GeV}$. 
ATLAS requires that the two photons are in the central region of the detector 
$| y_{\gamma} | < 2.37$, but  no 
photons  are allowed to be in the rapidity interval $1.37 < |y_\gamma|   < 1.56$.  However, when presenting 
the results 
of the measurements \cite{Aad:2014lwa}, 
the  ATLAS collaboration corrects for the second condition and, for this reason, 
we do not account for it in our  calculation.\footnote{We note that it is unclear to us why correcting 
for the missing rapidity region is a worthwhile thing to do since  such theoretical correction 
defies the original goal of comparing theoretical fiducial volume cross sections with the results 
of experimental {\it measurements}.}  It is required  that photons and jets are well-separated  
$\Delta R_{\gamma j} > 0.4$. Finally, we take 
the  branching ratio for the Higgs boson decay  to two photons to be ${\rm Br}(H \to \gamma \gamma ) = 2.35 \times 10^{-3}$.

\subsection{$H \to W^+ W^- \to e^+\mu^- \nu \bar \nu$}

As the second example, we consider $H+j$ production at the $13$~TeV LHC. The Higgs boson 
decays to $e^+\mu^- \nu \bar \nu$ final state through a pair of $W$ bosons.  
 To identify selection criteria, we apply 
kinematic cuts similar to those employed by the CMS collaboration in their studies of the $H \to W^+W^-$ production 
at the $8~{\rm TeV}$ LHC \cite{Chatrchyan:2013iaa}. We 
define jets using  the anti-$k_\perp$ algorithm with $\Delta R = 0.4$. Jets are required to have 
transverse momentum $p_{\perp, j} > 30~{\rm GeV}$ 
and be in the rapidity interval 
$-4.7 <y_{j}<4.7$. The harder of the two charged leptons must have transverse momentum 
$p_{\perp,l} > 20~{\rm GeV}$; the softer one  must have $p_{\perp,l} > 10~{\rm GeV}$.
The transverse missing energy in the event should exceed $E_{\perp, \rm miss} >20~{\rm GeV}$. 
Other cuts that we employ are i) 
the cut on the dilepton invariant mass  $m_{ll} > 12~{\rm GeV}$;
 ii) the cut on the transverse momentum of the dilepton pair $p_{\perp,ll} > 30~{\rm GeV}$ and iii)
the cut on the 
transverse mass  of the two $W$ bosons
$m_{\perp} = \sqrt{2 p_{\perp,ll} E_{\perp,{\rm miss}}(1-\cos \Delta\phi_{ll,{\rm miss}})} > 30~{\rm GeV}$.

\section{The results} 
\label{sect3}

\subsection{$ H \to \gamma \gamma$}

We consider production of the Higgs boson in association with a jet at the $8$~TeV LHC and 
compute the fiducial volume cross section and the kinematic distributions 
using the ATLAS selection criteria described 
in Section~\ref{sect2}. 
We begin with the fiducial volume cross section. For events that contain the Higgs boson 
and at least one jet, the ATLAS collaboration 
obtains~\cite{Aad:2014lwa}
\be
\sigma_{H+j}^{\rm fid}(8~{\rm TeV})  = 21.5 \pm 5.3({\rm stat.}) \pm ^{2.4}_{2.2}({\rm syst.}) \pm 0.6 ({\rm lumi}) \; {\rm fb}.
\label{eq0}
\ee
This result is significantly higher than what the   
fixed order computation predicts.  
For the $8~{\rm TeV}$ LHC we obtain inclusive fiducial  
$H(\gamma \gamma) +j$ production cross sections\footnote{We remind the reader that we work in the $m_t\to \infty$ limit. Moreover, we do not include subleading $qq$ channels beyond NLO, 
\emph{cf.} Refs.~\cite{hjetnnlo,hjetnnlo1}.}
\be
\sigma_{\rm LO}^{\rm fid} = 5.43^{+2.32}_{-1.49}~{\rm fb},
\;\;\; \sigma_{\rm NLO}^{\rm fid } = 7.98^{+1.76}_{-1.46}~{\rm fb},
\;\;\; \sigma_{\rm NNLO}^{\rm fid } = 9.45^{+0.58}_{-0.82}~{\rm fb}, 
\label{eq1}
\ee
where the central value corresponds to the factorization and renormalization scales set to a 
common value $\mu = m_H$ 
and the upper(lower) values to $\mu = m_H/2$ ($\mu = 2 m_H$), respectively. 
Theoretical  results indicate reasonable convergence of 
the perturbative expansion. Indeed,  for $\mu = m_H/2$  the 
NLO cross section is higher than the LO one by $26 \%$, and 
the NNLO cross section is higher than the NLO one  by  
only $3 \%$. The situation is worse but similar for $\mu = m_H$, where the NLO (NNLO) corrections amount to 
$47~(18)\%$, respectively.

\begin{figure}[t]
  \centering
  \includegraphics[width=0.49\textwidth]{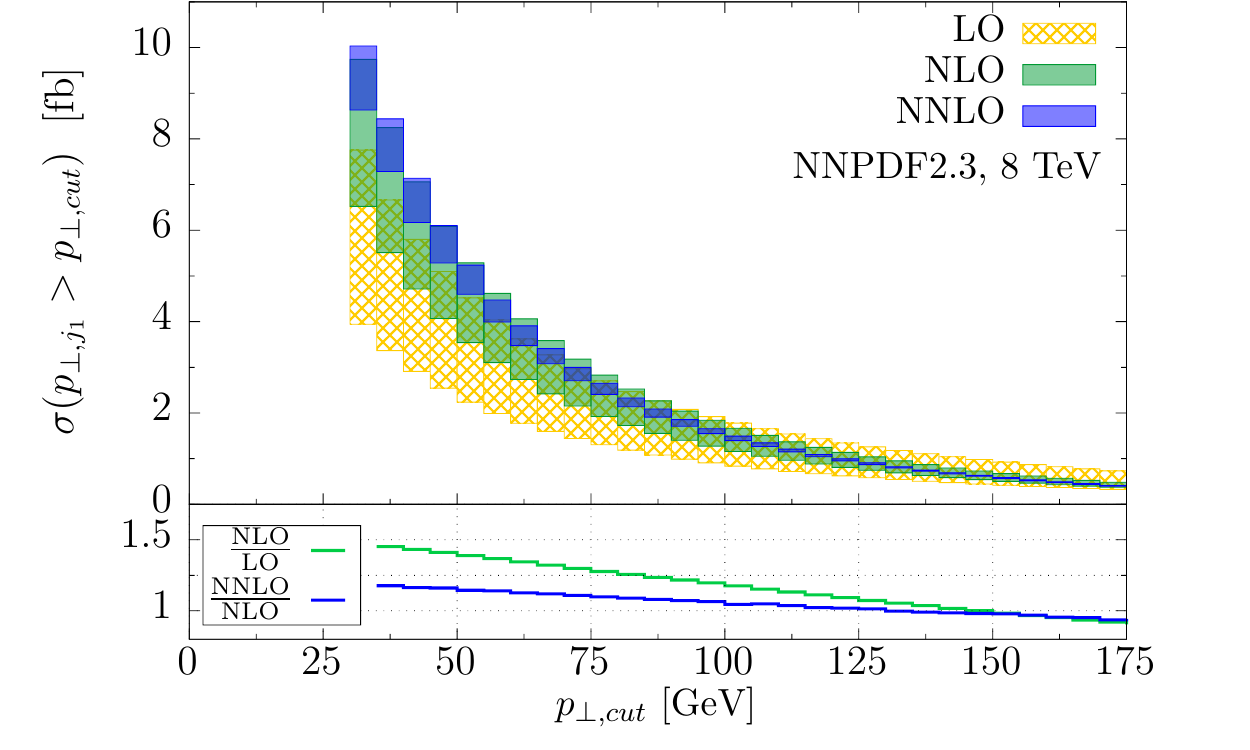}
  \includegraphics[width=0.49\textwidth]{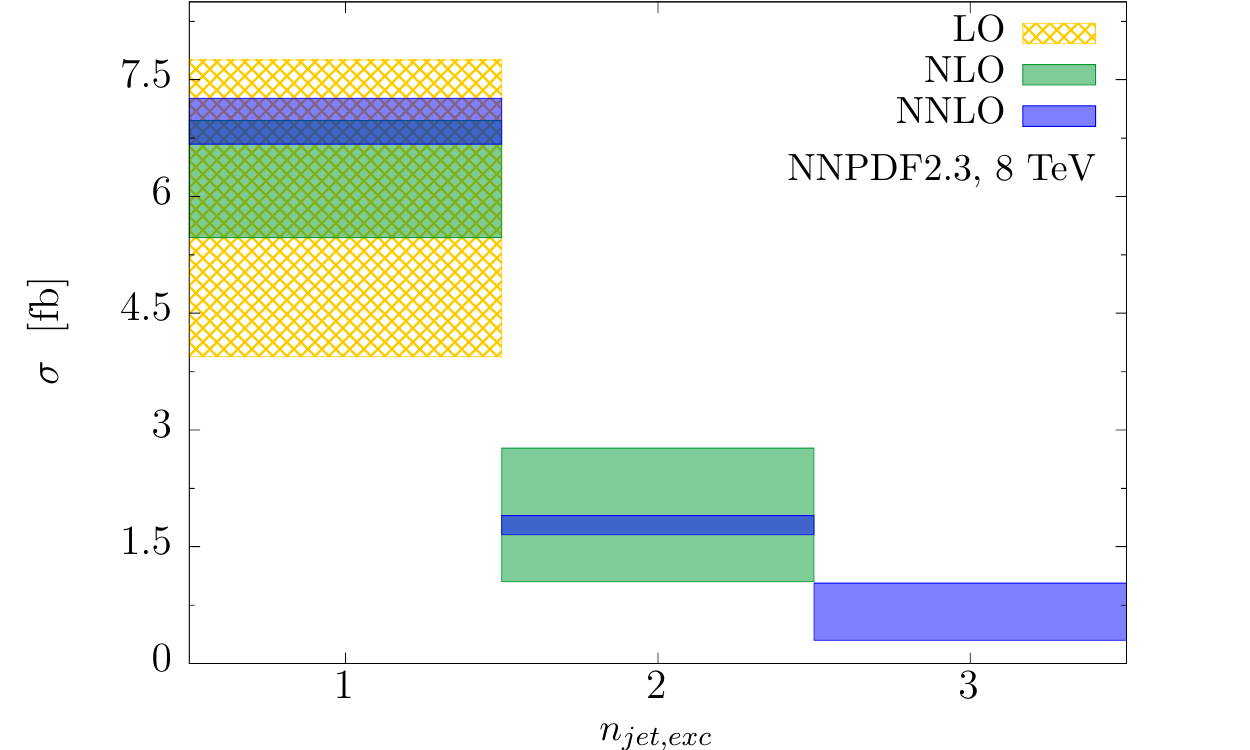}
  \caption{ Left pane: Fiducial cross section  for $pp \to H(\gamma \gamma) + j$  at $8$~TeV 
LHC as a function 
of the cut on the jet transverse momentum. The inset shows ratios of differential cross sections 
at different orders in perturbation theory for the factorization and the renormalization 
scales set to the mass of the Higgs boson.
Right  pane: fiducial cross sections for exclusive 
jet bins. The selection criteria are described in the text. 
}
  \label{fig1}
\end{figure}

It is interesting to point out that 
the quality of the perturbative expansion for 
$pp \to H+j$ appears to be somewhat better than for the inclusive 
Higgs boson production. As was pointed out earlier in the context of NLO computations, 
this feature may be related to a cancellation of Sudakov logarithmic corrections 
$ {\cal O}(\alpha_s^n \ln^{2n} p_{\perp, j}/m_H)$ 
that are present in the computation of the $H+j$ cross section, 
and other sources of large corrections that contribute to the inclusive rate.   
It was argued that this cancellation is accidental and for this 
reason  can not be expected to continue in higher orders of perturbation theory.  
Our  calculation questions this assertion. 
Indeed, it clearly demonstrates 
that, through  the NNLO in perturbative QCD,  
there is {\it no indication}  that  fixed order perturbation theory for $H+j$ 
production with the cut on the jet transverse momentum 
$p_{\perp, j} \ge 30~{\rm GeV}$   breaks down.

As a further   illustration of  this point, we show in the left  pane of Fig.~\ref{fig1} 
the cross sections for $pp \to H+j$ process as a function of  the jet transverse momentum 
cut $p_{\perp, \rm cut}$. It follows from this plot that both the NLO and the NNLO QCD corrections 
are moderate  for all values of $p_{\perp, \rm cut}$.
In particular,  the convergence of perturbative series 
for the cross section with $p_{\perp, \rm cut} = 30~{\rm GeV}$ 
does not appear to be  significantly worse 
than the convergence of  
perturbative predictions for larger values of the transverse momentum cut.

It is interesting to point out that in Ref.~\cite{Aad:2014lwa} a summary of the theoretical 
results for $H+j$ fiducial volume  cross sections was 
presented, based on exact NLO QCD computations 
and various   re-summations of potentially
enhanced terms \cite{minlo,bl, bm}. 
Compared to  these three results, our result is somewhat higher than 
the result of Ref.~\cite{minlo} and 
it is lower than the two results based on the 
calculations of Refs.\cite{bl,bm}.\footnote{Note however that finite
mass effects can increase the cross section by about $6\%$~\cite{Grazzini:2013mca,Banfi:2013eda}.} On the other 
hand, the uncertainty of our result is significantly smaller.

Nevertheless,  the difference between the measured $pp \to H+j$ cross section by the ATLAS collaboration 
 and our theoretical prediction is striking -- 
the central value measured by ATLAS  is higher than our results by a factor $2.1 - 2.5$, depending 
on the choice of the renormalization and factorization scale. 
If one accounts 
for the uncertainty of the experimental result, this  difference translates to, approximately, 
$2.4$ standard deviations and is not statistically significant. On the other 
hand, it is interesting to remark that the mismatch in the $H+j$ channel is stronger than 
the mismatch in the inclusive $pp \to H$ cross section, where the experimental cross section 
exceeds the theoretical one by  a factor of $1.4$~\cite{Aad:2014lwa}.
It will be interesting to watch  how results of  these  measurements 
will evolve in the future especially since, thanks to the availability 
of the NNLO QCD computation, the precision  of theoretical 
calculations  is quite  high.

\begin{figure}[t]
  \centering
  \includegraphics[width=0.49\textwidth]{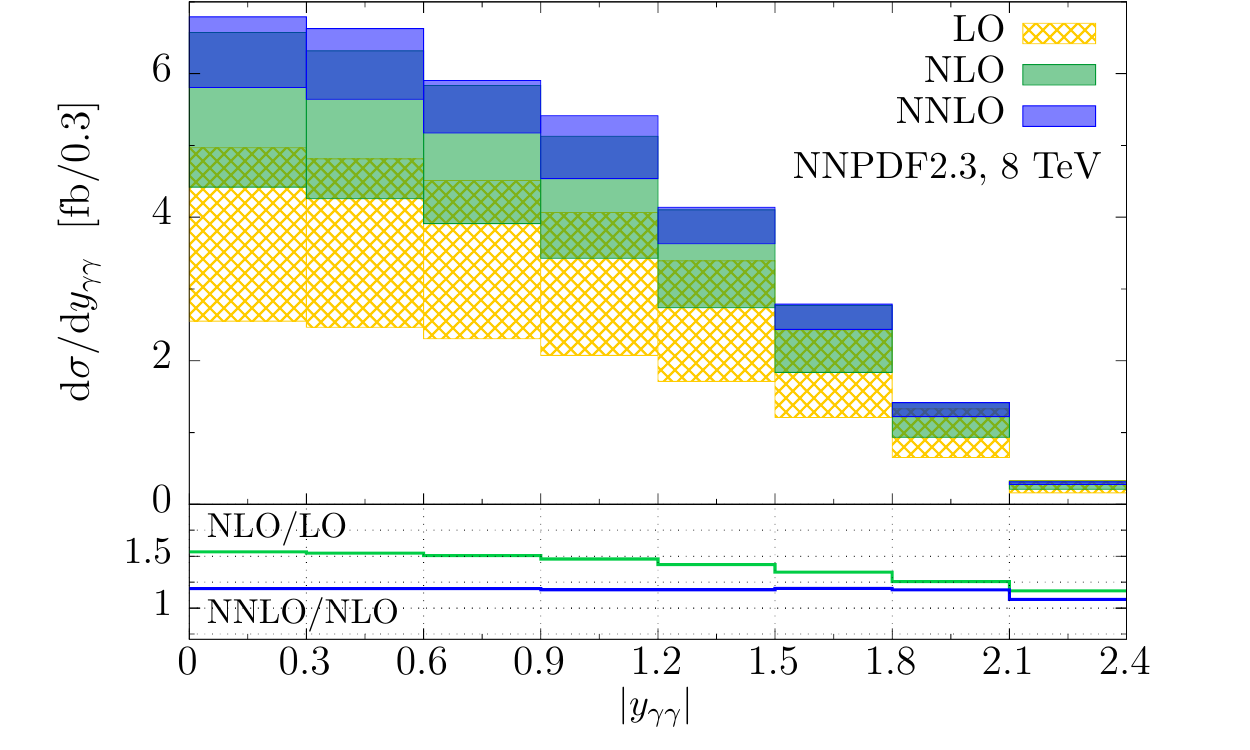}
  \includegraphics[width=0.49\textwidth]{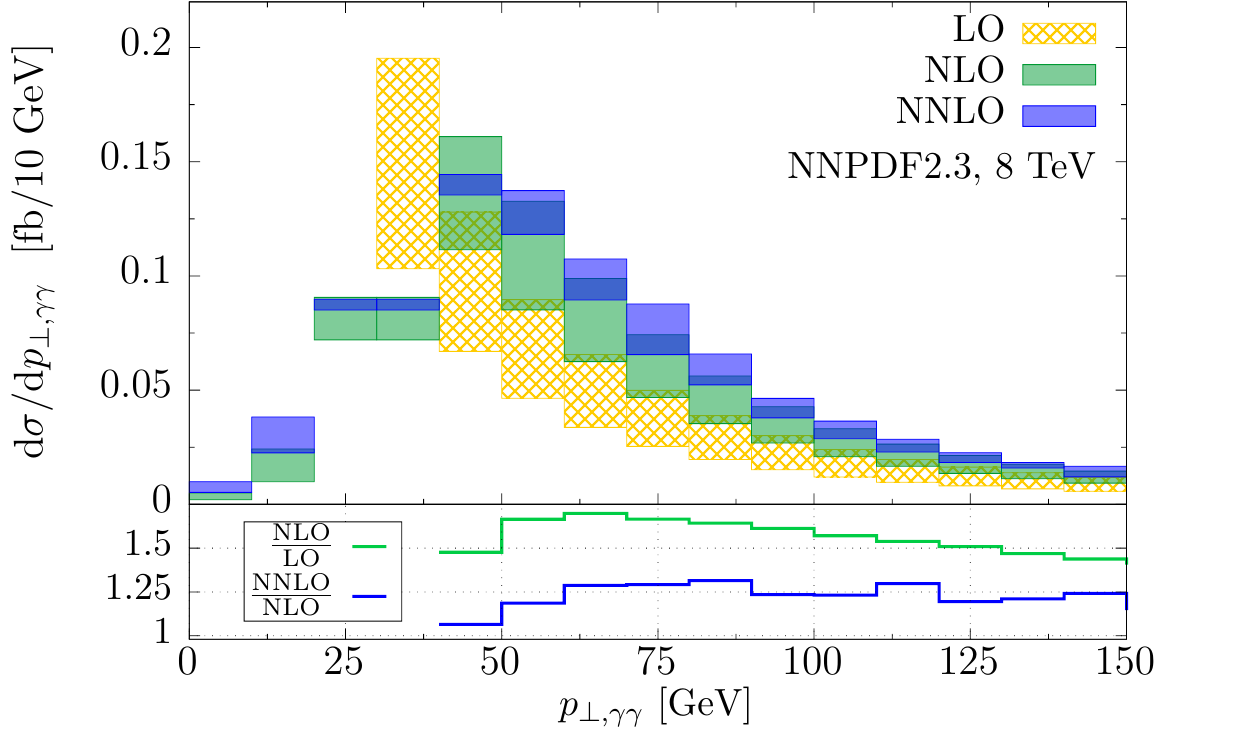}
  \caption{ Higgs boson rapidity (left) and transverse momentum (right) 
distributions  at the $8$ TeV LHC. The insets show ratios of differential cross sections 
at different orders in perturbation theory for the factorization and the renormalization 
scales set to the mass of the Higgs boson.
}
  \label{fig3}
\end{figure}

We can recast our results in  Eq.~(\ref{eq1})
into predictions for acceptances, at different orders in  perturbation theory. An acceptance 
is  defined as the ratio of a fiducial to total cross section  
$A = \sigma_{\rm fid}/\sigma$ for $H+j$ production.
When ratios of cross sections are computed, many sources of theoretical uncertainties cancel out and 
it is in 
general not possible 
to properly estimate  the uncertainty of the result by changing  factorization and renormalization scales within  
a prescribed interval. For this reason, it is useful  to know several orders in the 
perturbative expansion of  the acceptance, to estimate the precision with which it can actually be predicted. 
For the $8~{\rm TeV}$ LHC and the ATLAS setup, we find
\be
A_{\rm LO} = 0.594(4),\;\;\; A_{\rm NLO} = 0.614(3),\;\;\; A_{\rm NNLO} = 0.614(4). 
\label{eq3}
\ee
The perturbative expansion for the acceptances 
exhibits good convergence. Indeed, by comparing the central values, we find 
that the NLO acceptance is larger than the LO acceptance by $3$ percent, 
whereas there is no change going from NLO to NNLO.

Another interesting quantity is  the exclusive cross section for fixed number of jets. 
The corresponding results are shown in  the right pane of Fig.~\ref{fig1}.   We observe 
good convergence of the perturbative expansion for  the exclusive $H+j$ and  $H+2j$  production
cross sections  at $8~{\rm TeV}$ LHC.  We can not discuss the perturbative  behavior of the $H+3j$ cross section 
since it enters our computation only at   leading order in perturbative QCD. 

\begin{figure}[t]
  \centering
  \includegraphics[width=0.49\textwidth]{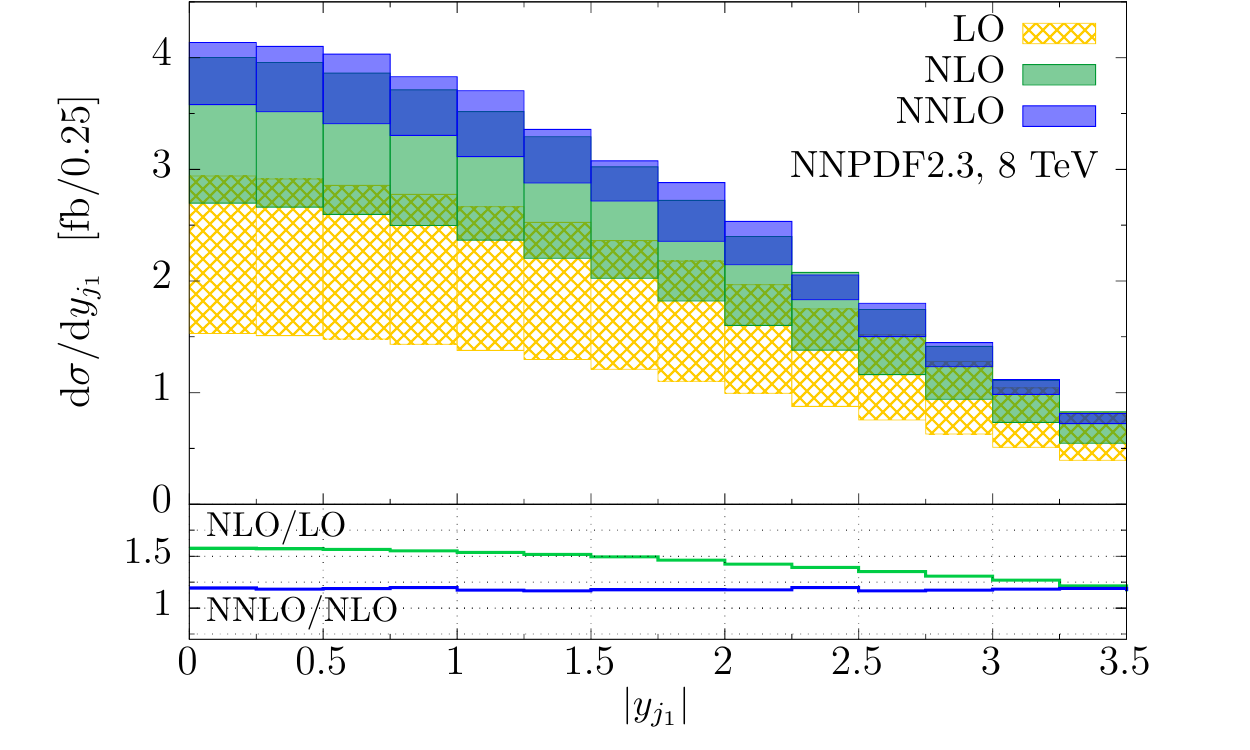}
  \includegraphics[width=0.49\textwidth]{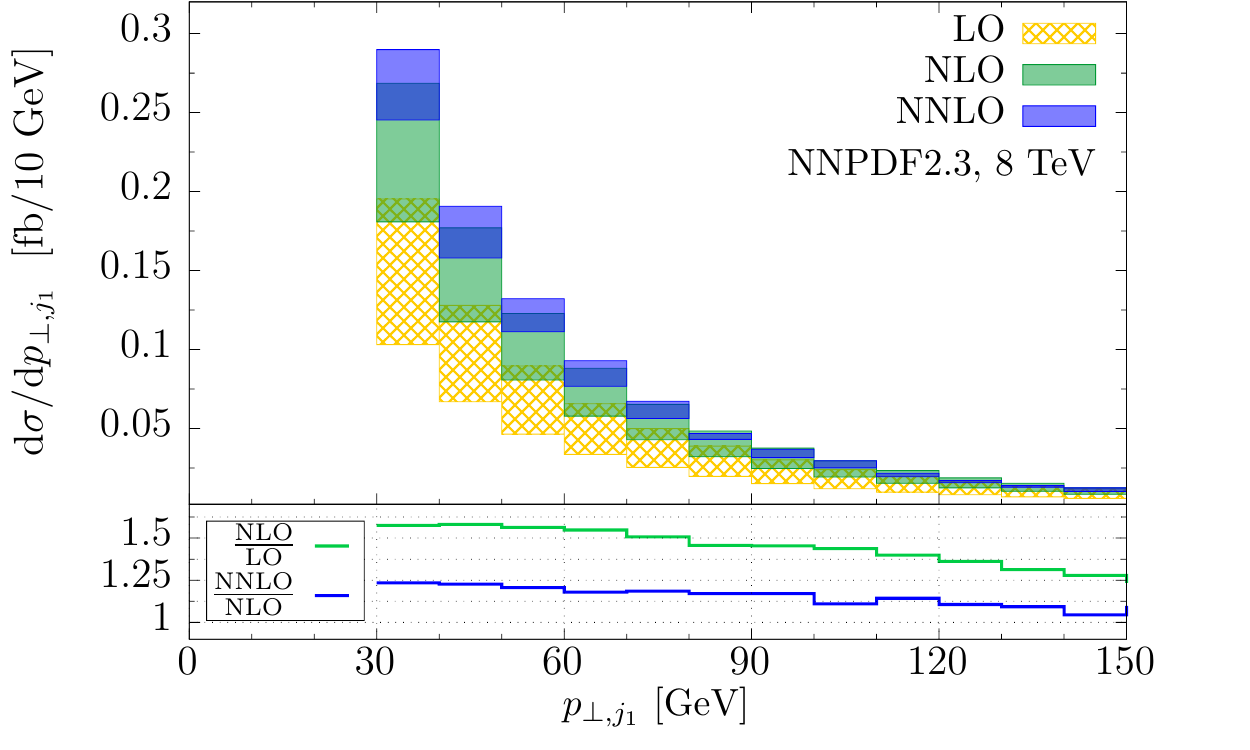} 
  \caption{ Rapidity and transverse momentum distributions of the most energetic jet
 at the $8$~TeV LHC. The insets show ratios of differential cross sections 
at different orders in perturbation theory for the factorization and the renormalization 
scales set to the mass of the Higgs boson.
}
  \label{fig4}
\end{figure}

We now turn  to  kinematic distributions studied by the ATLAS collaboration. 
They can be divided into three 
categories:   transverse momentum and rapidity distributions of the Higgs boson;
transverse momentum, rapidity and the transverse energy  
distributions of the accompanying QCD radiation 
 and, finally,  kinematic distributions of individual photons. The latter includes the transverse momentum  and  the 
rapidity distributions as well as the distribution of the photon decay angle in the Collins-Soper reference frame. 
We can compute all these kinematic distributions through NNLO in perturbative 
QCD, using exactly the same setup  that the ATLAS collaboration employs  
in the actual measurement.  

We begin with the discussion of the rapidity and the transverse momentum distributions of the Higgs 
boson in events with at least one jet, see Fig.~\ref{fig3}.  
The pattern of radiative corrections is similar to the 
fiducial cross section case that we just discussed. In the two  plots in Fig.~\ref{fig3}
the relative magnitude  of radiative corrections  is illustrated 
in lower panes, where 
ratios of NLO to LO and NNLO to NLO distributions  at $\mu = m_H$ are displayed. 
We will  refer to such ratios as $K$-factors.  We note that similar 
to the case of the inclusive Higgs  boson  production $pp \to H$, the NNLO enhancement of the Higgs 
boson rapidity distribution in $pp \to H+j$  process is independent 
of the rapidity.  On the contrary, 
the $K$-factors for transverse momenta distributions have a more interesting shape. Indeed,  we observe 
the instability of ${\rm d}\sigma/{\rm d} p_{\perp, H}$ at the value of the 
Higgs boson transverse momentum equal to the 
value of the jet transverse momentum cut. This is the 
manifestation  of the so called 
Sudakov-shoulder effect~\cite{Catani:1997xc}.  Just above  $p_{\perp, H} \sim 30~{\rm GeV}$,
the NNLO corrections  are small but they increase to about $30\%$ at around $p_{\perp, H} \sim 75~{\rm GeV}$ 
and then start to decrease again. 

\begin{figure}[t]
  \centering
  \includegraphics[width=0.49\textwidth]{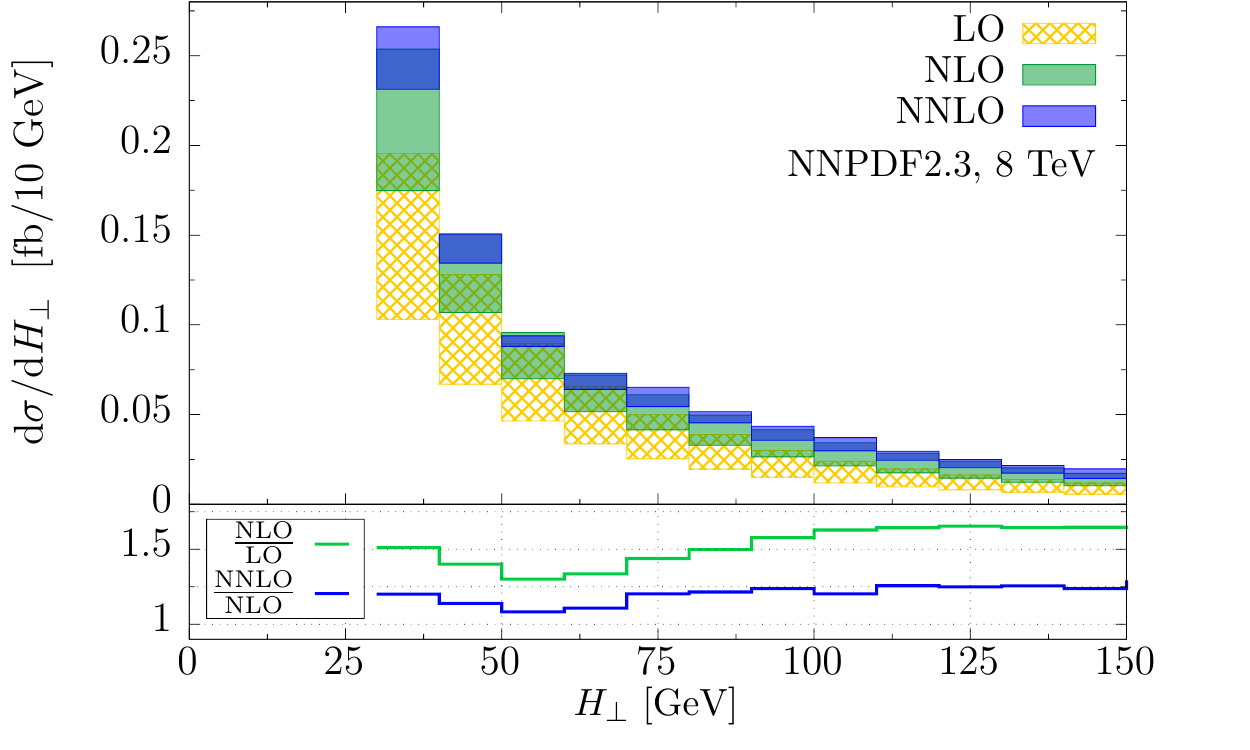} 
  \caption{ The distribution of the total transverse energy 
$H_\perp$ in $H+j\to\gamma\gamma +j$ production for the 8 TeV LHC. 
The inset shows ratios of differential cross sections 
at different orders in perturbation theory for the factorization and the renormalization 
scales set to the mass of the Higgs boson.
}
  \label{fig5}
\end{figure}

Next, we consider kinematic distributions of  the QCD radiation that accompanies the Higgs boson 
production.  The rapidity and the transverse momentum distributions 
of the hardest jet are shown in Fig.~\ref{fig4}. 
Similar to the QCD corrections to the Higgs boson rapidity distribution, the NNLO $K$-factor 
for the hardest jet rapidity distribution is flat. 
The transverse momentum distribution is re-shaped slightly, with corrections being  larger at smaller 
$p_{\perp,j}$ and smaller at  larger $p_{\perp,j}$.  
The distribution of  the total transverse energy $H_\perp$ of QCD radiation 
defined as  $H_\perp = \sum \limits_{i}^{} p_{\perp, j_i}$ is shown in Fig.~\ref{fig5}.  The sum is taken  over all jets 
{\it observed} in a given events.  The NNLO QCD corrections for this observable are smaller for smaller $H_\perp$,  but, eventually, they  increase and flatten out.  This is similar to what happens to this 
observable already at next-to-leading order.

\begin{figure}[t]
  \centering
  \includegraphics[width=0.49\textwidth]{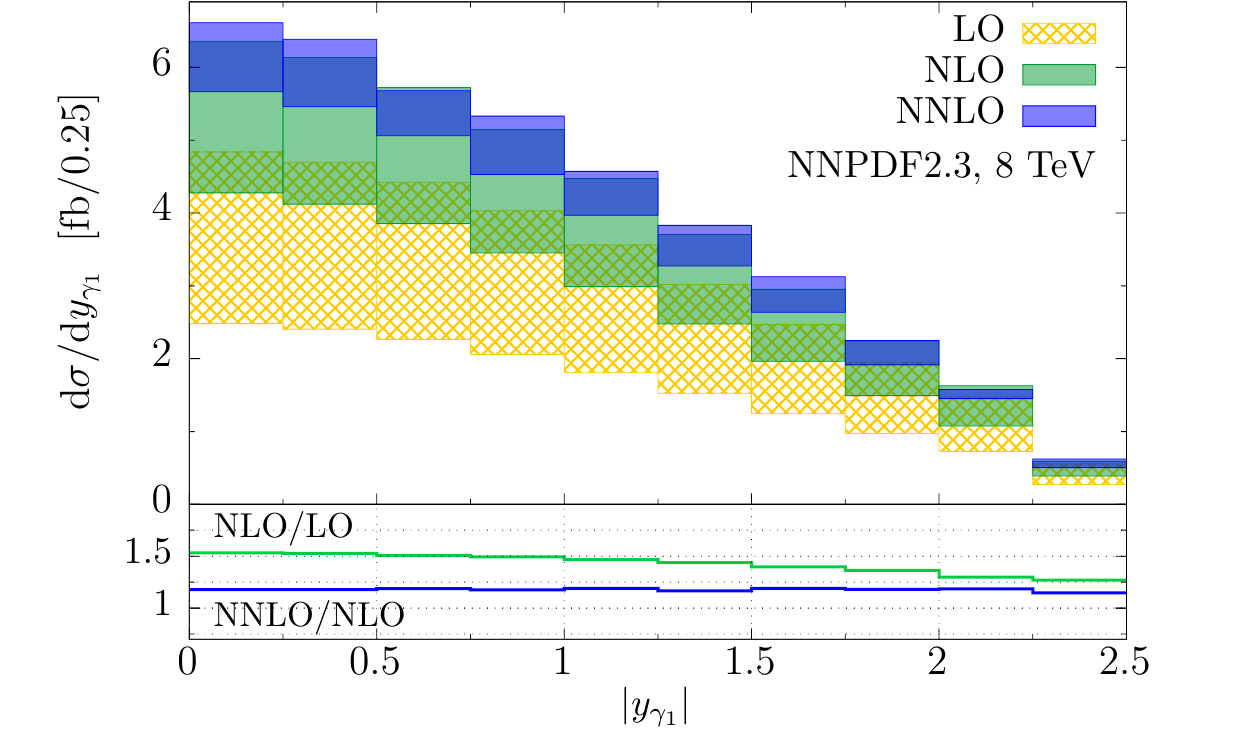}
  \includegraphics[width=0.49\textwidth]{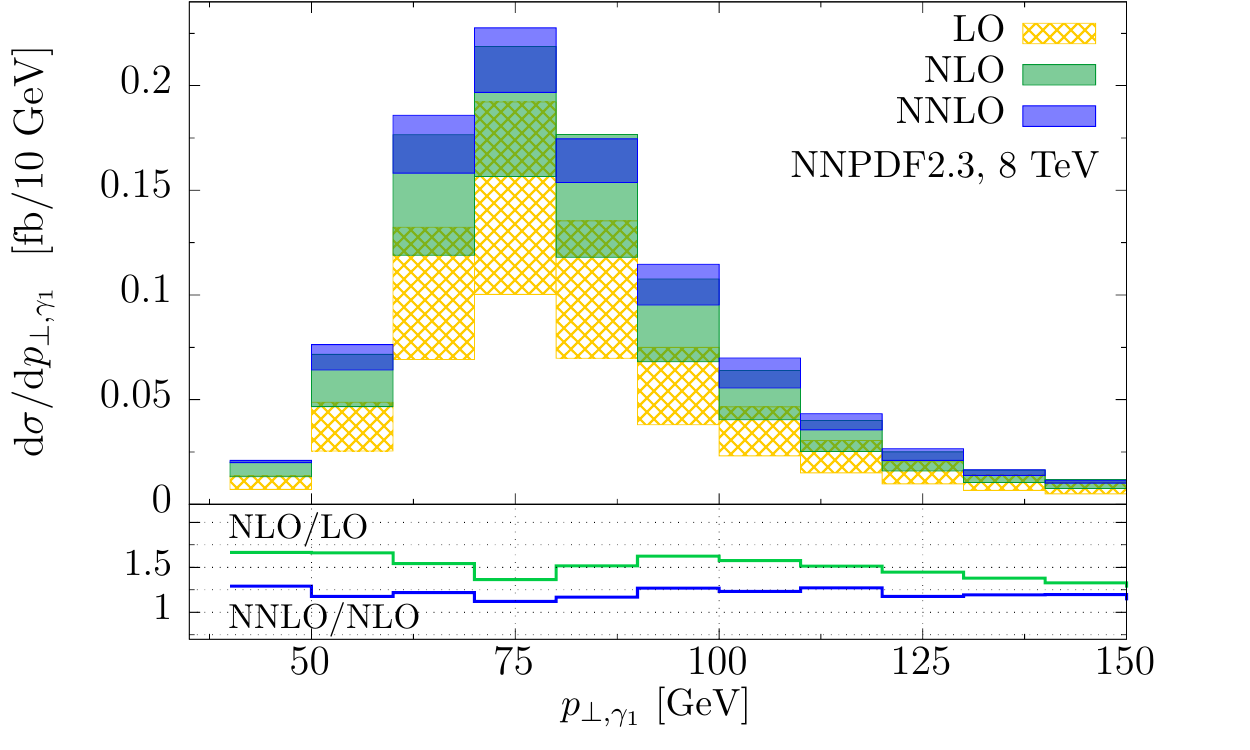} 
  \includegraphics[width=0.49\textwidth]{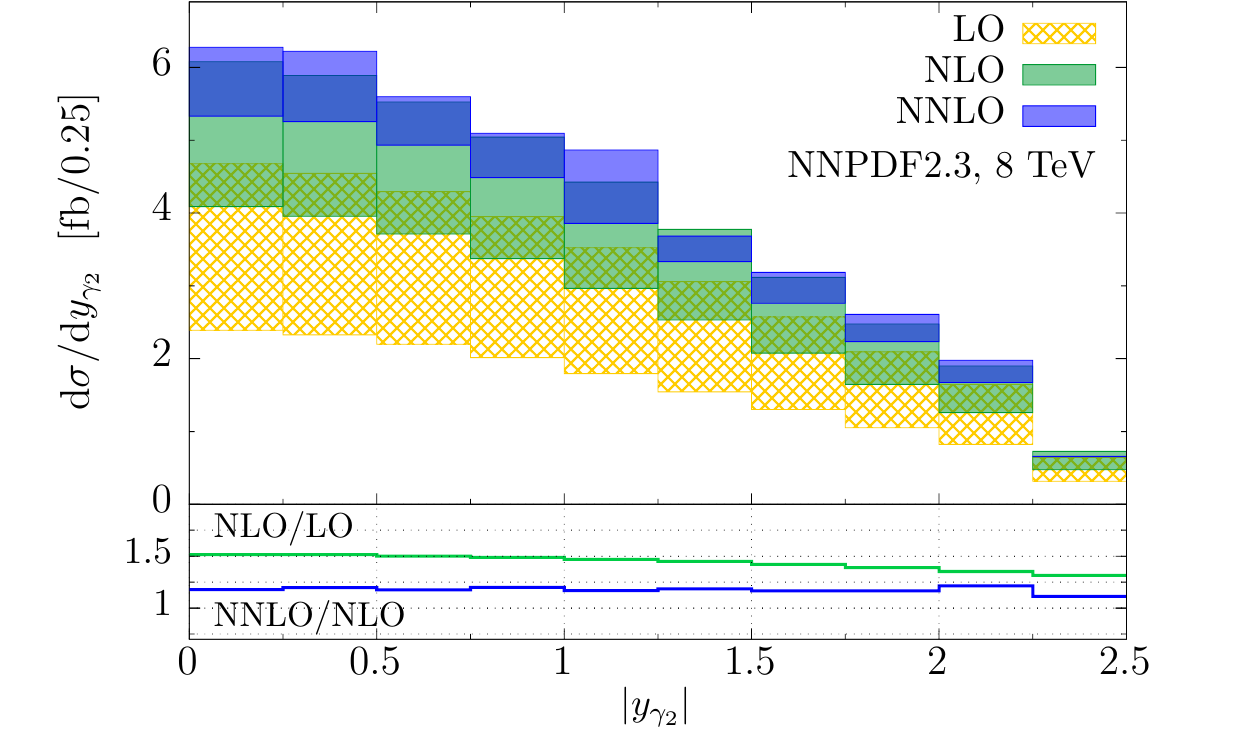}
  \includegraphics[width=0.49\textwidth]{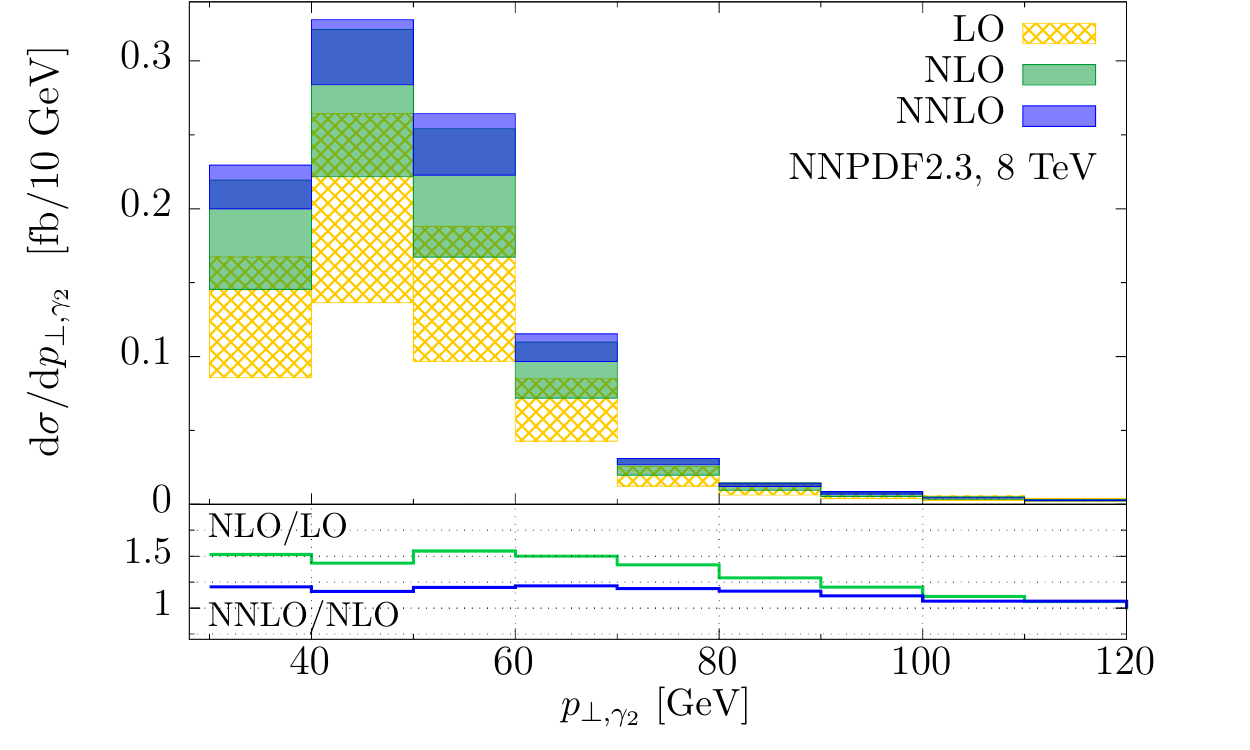} 
  \caption{ Rapidity and transverse momentum distributions of the harder and softer photons  
in $pp \to H + j$ at the $8$ TeV LHC. 
The insets show  ratios of differential cross sections 
at different orders in perturbation theory for the factorization and the renormalization 
scales set to the mass of the Higgs boson.
}
  \label{fig6}
\end{figure}

Finally, we turn to distributions that describe kinematic properties of individual photons that originate from the Higgs boson 
decays.   In Fig.~\ref{fig6}, we show transverse momentum and rapidity distributions of the two photons.
Similar to other cases, we find a uniform enhancement of the rapidity distribution and some shape-dependent NNLO 
effects in transverse momenta distributions. However, 
the shape-dependence of QCD corrections is significantly reduced 
at NNLO compared to the NLO case. This is 
particularly true for events with transverse momenta at around maximal values 
for the transverse momenta of both the harder and the softer photons.

\begin{figure}[t]
  \centering
  \includegraphics[width=0.49\textwidth]{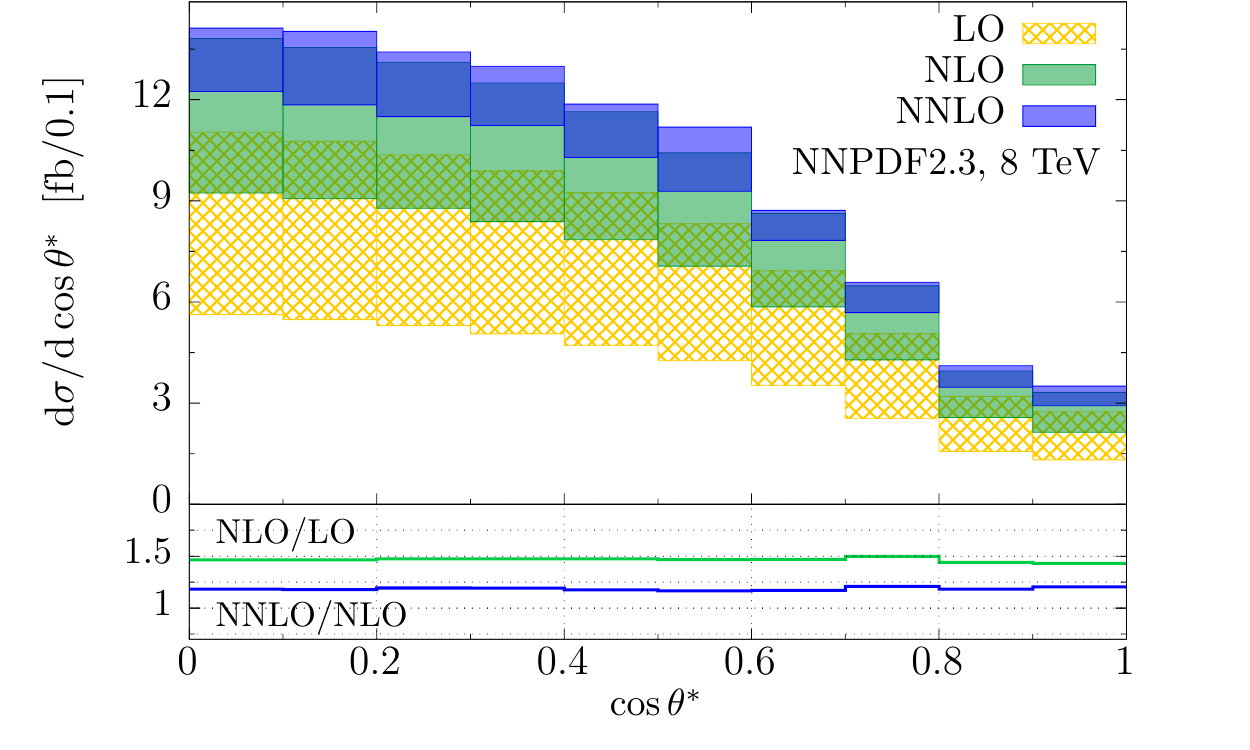}
  \caption{ Distribution of the photon decay angle in the Collins-Soper frame 
 at the $8$ TeV LHC. The inset shows ratios of differential cross sections 
at different orders in perturbation theory for the factorization and the renormalization 
scales set to the mass of the Higgs boson.
}
  \label{fig7}
\end{figure}

In Fig.~\ref{fig7}, we show the 
distribution of the photon decay angles in the Collins-Soper reference frame 
defined as 
\be
\cos \theta^* = |\sinh ( y_{\gamma_1} - y_{\gamma_2} ) | \frac{2 p_{\perp \gamma_1} p_{\perp \gamma_2} }{
m_H^2 \sqrt{1 + (p_{\perp,H}/m_H)^2} }.
\ee
The $\cos \theta^*$ distribution  is important for studying  the spin-parity quantum numbers of the Higgs boson~\cite{Aad:2013xqa}. 
We find that the shape of this distribution is very well predicted by leading order 
perturbative QCD computations, 
with both NLO and NNLO QCD corrections providing a uniform enhancement.  
This observation should enable the reduction  of the uncertainty associated 
with the  modelling of this observable and, perhaps, lead  to  improved limits on exotic features 
of the observed Higgs resonance.

\begin{figure}[t]
  \centering
  \includegraphics[width=0.49\textwidth]{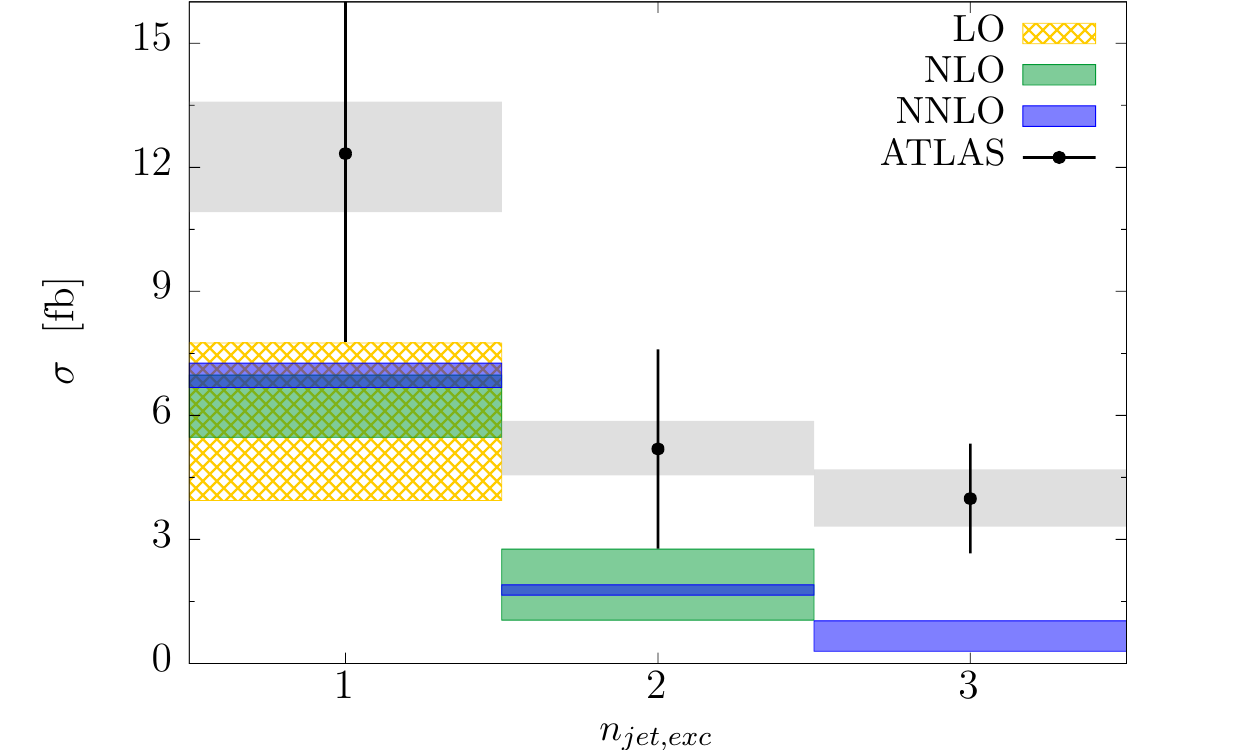}
  \includegraphics[width=0.49\textwidth]{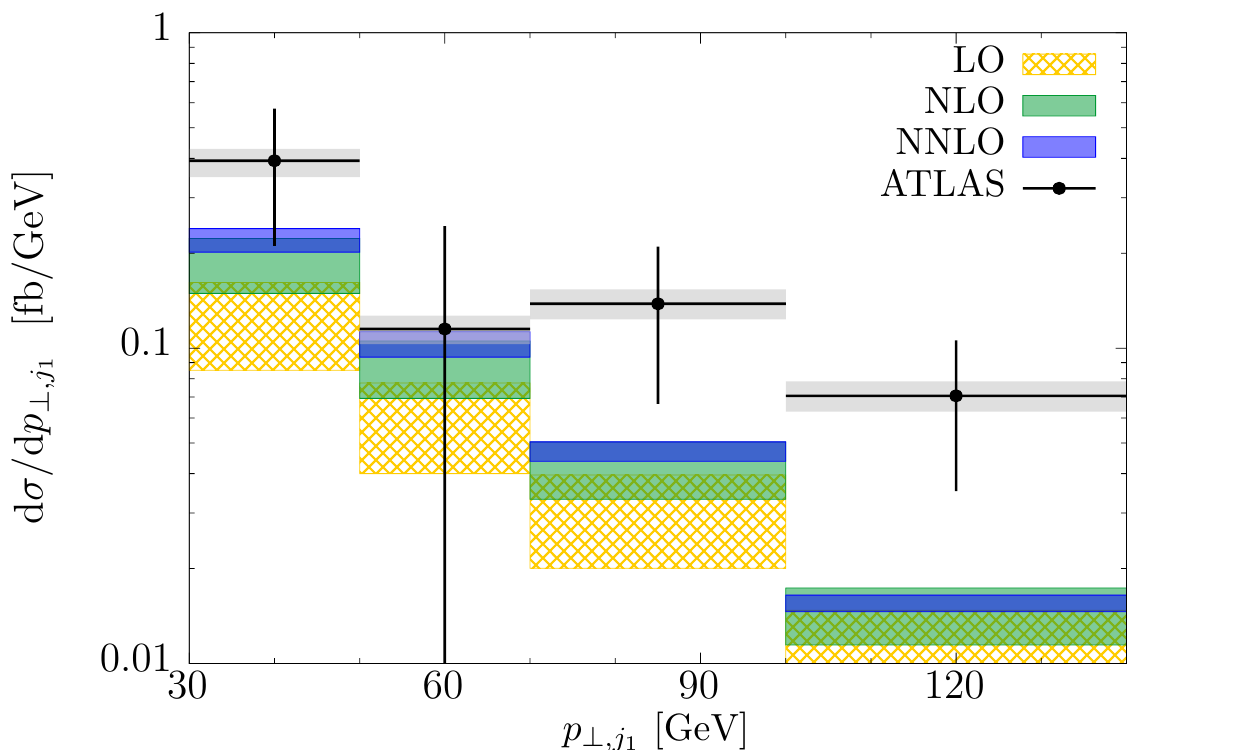}
  \caption{ 
Left  pane: comparison of exclusive jet cross sections in $pp \to H+j \to \gamma \gamma + j$ 
computed in this paper and measured by the ATLAS  collaboration.  Right pane: comparison 
of the leading jet transverse momentum distribution. 
The selection criteria are described  in the text. 
}
  \label{fig8a}
\end{figure}

Finally, in Fig.~\ref{fig8a}  we compare  the 
ATLAS measurements with our computations of the fiducial 
volume signal in $pp \to H+j \to \gamma \gamma + j$. 
The  inclusive  one-jet cross section was already discussed at the beginning of this Section; 
we remind the reader that the result of the 
ATLAS measurement is significantly higher than the NNLO 
QCD prediction for the inclusive one-jet  cross section. 
In the left pane of Fig.~\ref{fig8a} we present a similar comparison for the exclusive jet 
cross sections. We see that the situation is similar for all jet multiplicities and that the 
discrepancy increases for higher-multiplicity bins. In the right 
pane of Fig.~\ref{fig8a}, 
theoretical and experimental results for the  
transverse momentum distribution  of the hardest 
 jet are compared. For this observable, 
the ATLAS results are higher than the theoretical predictions in  
all $p_\perp$-bins except one where the 
experimental error is the largest. It is also clear that the shapes 
of theoretical and experimental 
distributions  are different. 
It follows from the plots in Fig.~\ref{fig8a} that, currently, 
 the ATLAS data is not precise enough to allow for  a meaningful comparison 
with available theoretical predictions.  This will 
undoubtedly change once enough luminosity at the 13 TeV LHC is collected.

\subsection{$ H \to W^+W^-\to e^+\mu^-\nu\bar\nu$}

In this subsection, we present the results for  
the process $pp \to H+j \to W^+W^- + j$ at the $13~{\rm TeV}$ LHC. 
The selection criteria are described in Section~\ref{sect2}. In principle, many kinematic 
distributions that can be studied in the $H+j$ production process are independent of the decay 
mode of the Higgs boson. To avoid overlap with the previous subsection, we present here 
only those distributions that are particular to the $W^+W^-$ final state. 

\begin{figure}[t]
  \centering
  \includegraphics[width=0.49\textwidth]{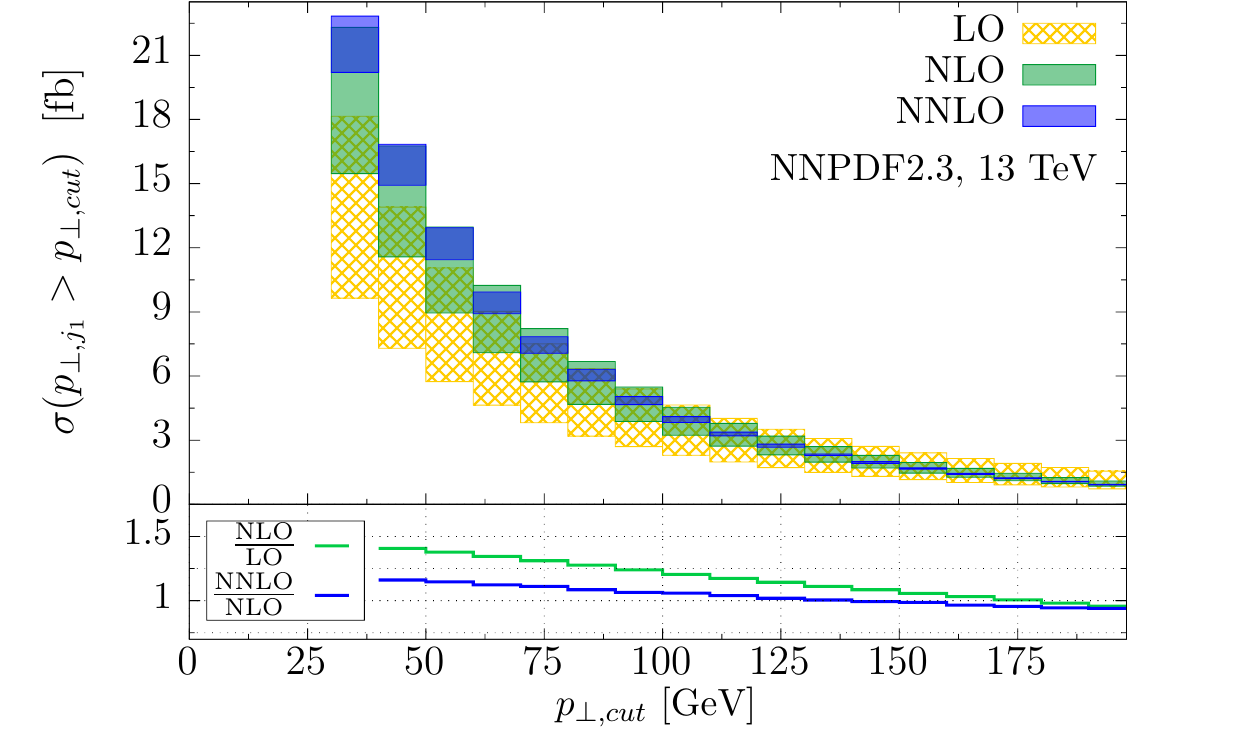}
  \includegraphics[width=0.49\textwidth]{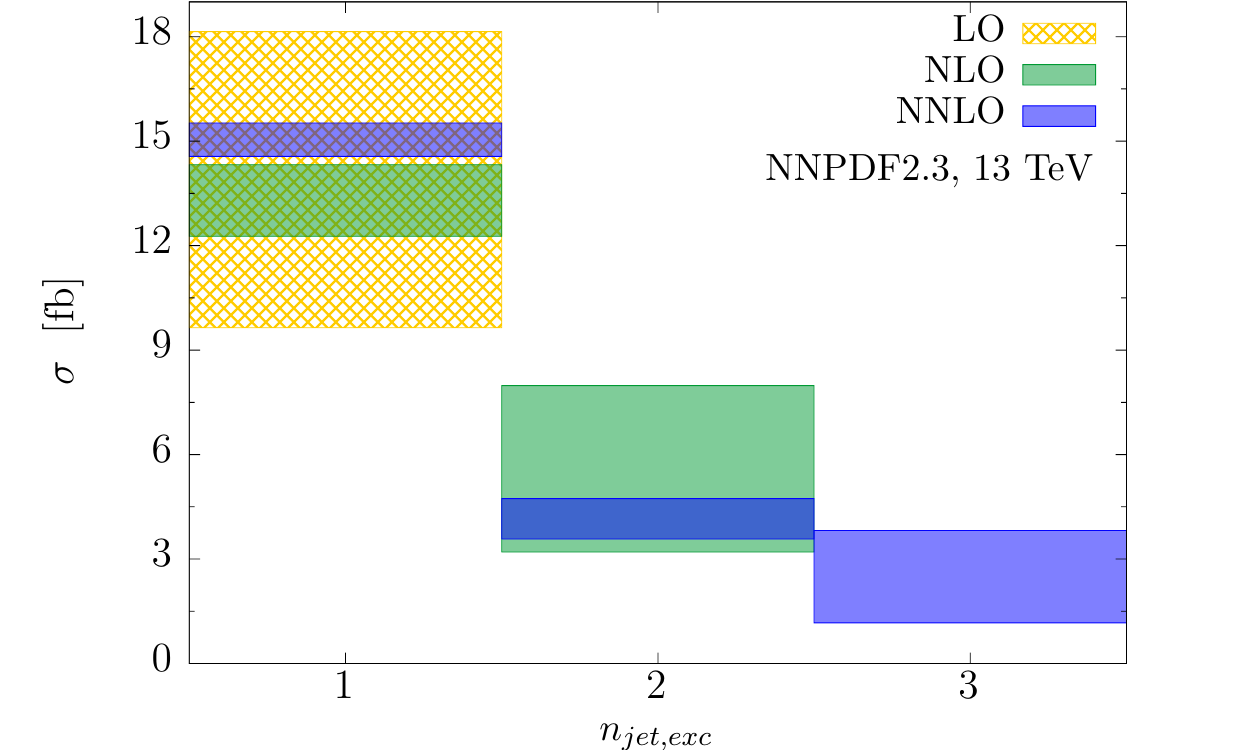}
  \caption{ 
Left pane: the production cross section for $pp \to H + j \to W^+W^- + j \to e^+\mu^-\nu \bar \nu +j$ at the 
$13~{\rm TeV}$ LHC is shown 
as a function of the jet transverse momentum cut. 
The inset shows ratios of differential cross sections 
at different orders in perturbation theory for the factorization and the renormalization 
scales set to the mass of the Higgs boson.
Right  pane: exclusive jet cross sections for 
$pp \to H(e^+\mu^-\nu \bar \nu) +j$ at the $13~{\rm TeV}$ LHC.  The selection criteria are described 
in the text. 
}
  \label{fig8}
\end{figure}

We begin, however, with the discussion of the fiducial  cross sections. We find 
\be
\sigma^{\rm fid}_{\rm LO} = 13.0^{+5.1}_{-3.4}~{\rm fb}, 
\;\;\;
\sigma^{\rm fid}_{\rm NLO} = 18.6^{+3.7}_{-3.1}~{\rm fb}, 
\;\;\;
\sigma^{\rm fid}_{\rm NNLO} = 21.9^{+0.9}_{-1.7}~{\rm fb}. 
\ee
In general, the perturbative expansion of the 
$13$ TeV cross sections is similar to what was observed 
at  $8$ TeV.  At $\mu = m_H$, the NLO cross section is larger than the LO cross section by $43\%$ 
and the NNLO cross section exceeds the NLO cross section by $18\%$. 
In Fig.~\ref{fig8} we display results for cross sections as a function of the 
jet transverse momentum cut and exclusive jet cross sections at different orders in perturbation theory.
The behavior of the exclusive one-jet cross section at $13~{\rm TeV}$ is slightly worse than 
at $8~{\rm TeV}$. We attribute this to a too small scale variation at 
NLO, 
which leads to the NNLO result for the one-jet 
cross section being  outside of the 
NLO scale variation band.

\begin{figure}[t]
  \centering
  \includegraphics[width=0.49\textwidth]{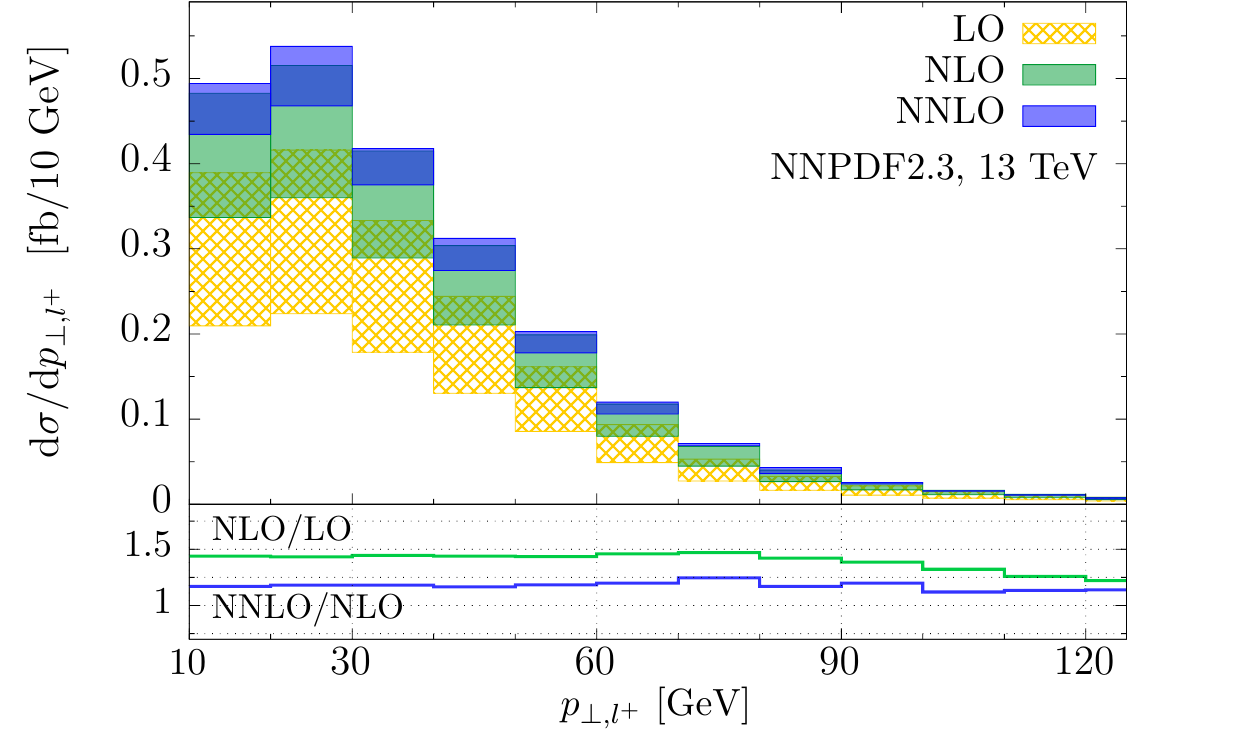}
  \includegraphics[width=0.49\textwidth]{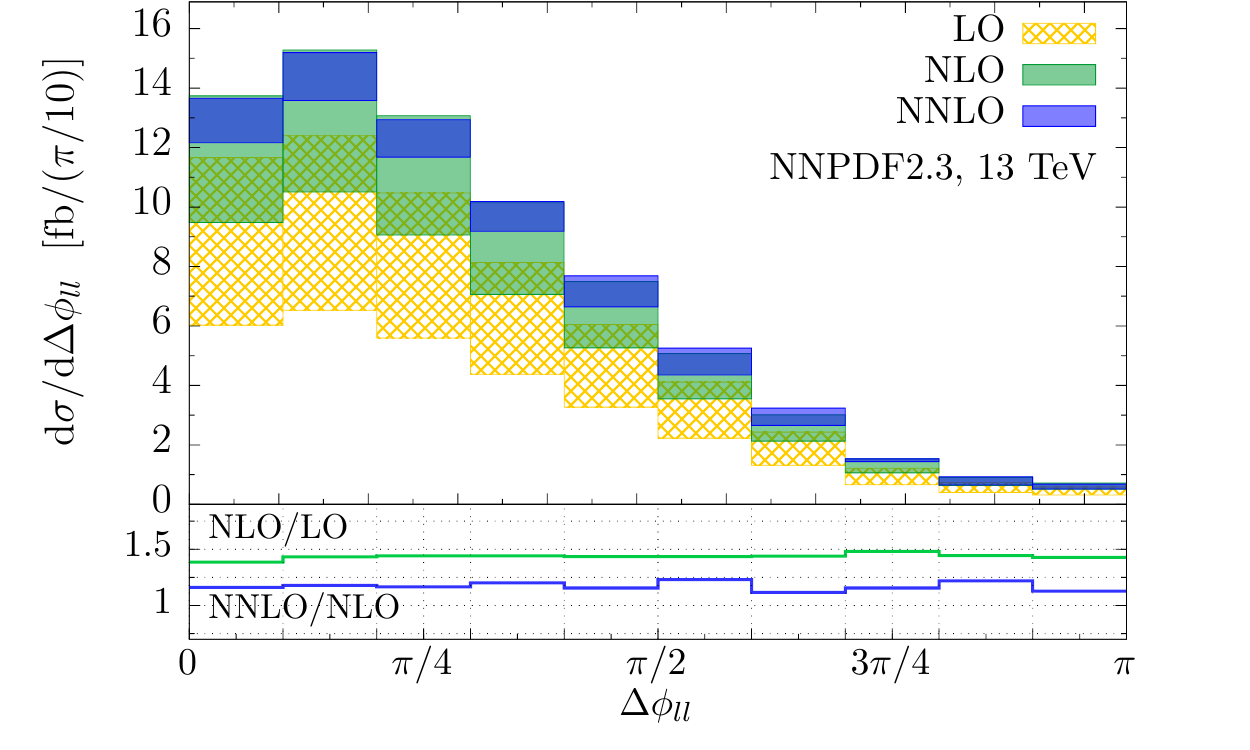}
  \caption{ 
Left pane: the transverse momentum distribution of a positively charged 
lepton in $pp \to H(e^+\mu^-\nu\bar\nu) + j$ at the 
$13~{\rm TeV}$ LHC. Right  pane: the distribution of the azimuthal opening angle of the 
two leptons  in 
$pp \to H(e^+\mu^-\nu\bar\nu) + j$ 
at the $13~{\rm TeV}$ LHC.  The selection criteria are described 
in the text. The insets show ratios of differential cross sections 
at different orders in perturbation theory for the factorization and the renormalization 
scales set to the mass of the Higgs boson.
}
  \label{fig9}
\end{figure}

\begin{figure}[t]
  \centering
  \includegraphics[width=0.49\textwidth]{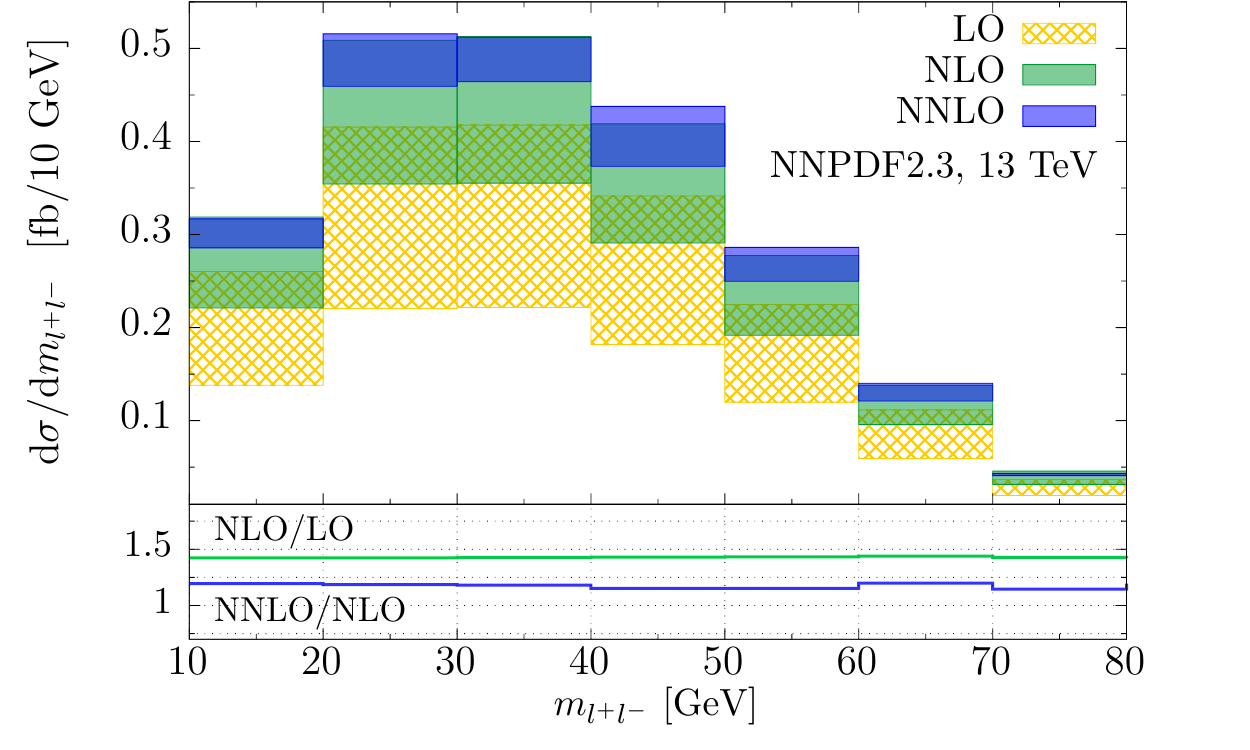}
  \includegraphics[width=0.49\textwidth]{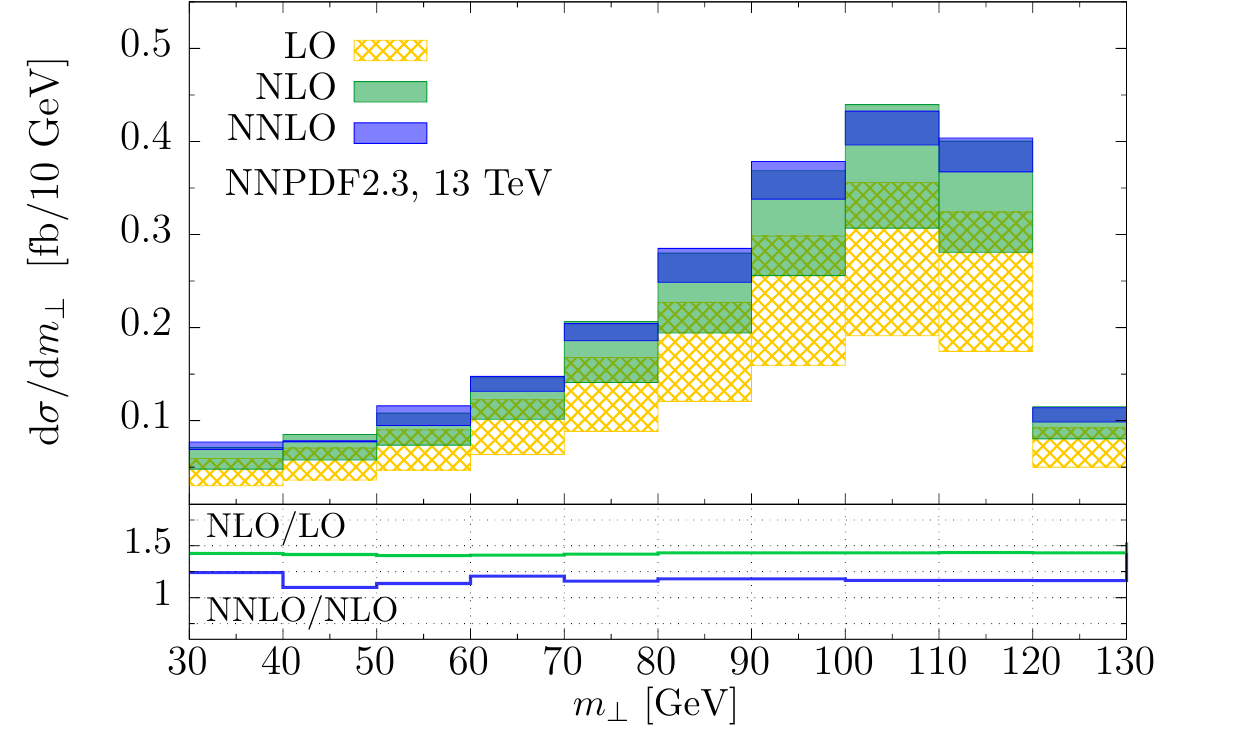}
  \caption{ 
Left pane: the lepton invariant mass  distribution in $pp \to H(e^+\mu^-\nu \bar \nu) +j$ at the 
$13~{\rm TeV}$ LHC. Right  pane: the $W^+W^-$ boson transverse mass distribution in 
$pp \to H(e^+\mu^-\nu \bar \nu) +j$ at the $13~{\rm TeV}$ LHC.  The selection criteria are described 
in the text. The insets show ratios of differential cross sections 
at different orders in perturbation theory for the factorization and the renormalization 
scales set to the mass of the Higgs boson. 
}
  \label{fig10}
\end{figure}

Selection criteria for the Higgs signal in $H \to W^+W^-$  as well as analysis 
of anomalous couplings in this process rely on  kinematic distributions of charged leptons.
A good understanding of these distributions is therefore important. In Fig.~\ref{fig9} 
we show the transverse momentum distribution of a positively charged 
lepton in $pp \to H + j \to W^+W^- + j \to e^+\mu^-\nu \bar \nu +j$ at the 
$13~{\rm TeV}$ LHC and the azimuthal opening angle distribution of the 
two leptons. In both cases, QCD radiative corrections do not change the shapes of the distributions 
significantly. The distribution of the invariant masses of the two leptons $m_{l^+l^-} $
and the transverse mass  $m_\perp$ are displayed in Fig.~\ref{fig10}; the NNLO QCD corrections 
to those distributions are remarkably uniform.

\section{Conclusions} 
\label{conc}

We  extended the recent NNLO QCD computation of the 
$H+j$ production in proton collisions by including  decays of the Higgs 
bosons to electroweak gauge bosons $H \to \gamma \gamma$, $H \to W^+W^-$ and $H \to ZZ$. 
Leptonic  decays of $Z$'s and $W$'s, with all spin correlations, are fully accounted for. 
This allows us to calculate 
 fiducial volume cross sections and various kinematic distributions through 
NNLO in perturbative QCD in a manner that is fully consistent with selection criteria applied 
in experiments. In particular, it becomes possible -- for the first time -- to confront fiducial 
volume studies   of the $pp \to H+j \to \gamma \gamma +j$ process
 performed by the ATLAS collaboration at the 
$8$~TeV LHC~\cite{Aad:2014lwa} with NNLO QCD predictions.

We presented a number of results for fiducial volume cross sections, 
acceptances  and various kinematic distributions for both inclusive and exclusive 
$H(\gamma \gamma) +j$ production processes. We find no indication that 
perturbative QCD breaks down and  requires resummation for the jet cut as low 
as $30~{\rm GeV}$.  We have also studied the $WW$ final state in $H+j$ production 
at the $13~{\rm TeV}$ LHC. We find that most of the kinematic distributions 
used to distinguish  this channel from backgrounds, 
show uniform enhancement when NNLO QCD corrections are included.  Changes of shapes of such 
distributions -- if any -- are already properly captured by the 
NLO QCD computations. 

As a final remark, we note that the availability of higher order QCD predictions for fiducial volume 
quantities should allow direct and precise studies of the ratios of 
Higgs signals.  The idea that ratios of cross sections are useful for 
reducing theoretical uncertainties is, of course, well-known and 
appreciated.  However, given the availability of the NNLO QCD computations for fiducial cross sections, 
no extrapolations should be required. As an illustration
we compute the ratio of fiducial cross sections 
for $pp \to H + j \to \gamma \gamma +j$ 
at the $8$ TeV LHC and the 
$pp \to H + j \to WW^* +j \to e^+\mu^- \nu \bar \nu + j$ at the $13$ TeV LHC.
We obtain  
\be
R_{WW/\gamma \gamma}  = \frac{\sigma_{H+j}^{WW \to e^+\mu^- \nu \bar \nu, 13 ~\rm TeV }}{\sigma_{H+j}^{\gamma \gamma, 8~\rm TeV}}
  = 2.39^{-0.06}_{+0.04},\;\;\;2.33^{-0.04}_{+0.05},\;\;\;2.32^{-0.04}_{+0.02},
\ee
at leading, next-to-leading and next-to-next-to-leading order in perturbative QCD, respectively. 
The convergence of the series is striking;  at NNLO QCD, we are able to predict $R_{WW/\gamma \gamma}$
with the precision of better than $2\%$.  Since 
$R_{WW/\gamma \gamma}$ is proportional 
to the ratio of the Higgs couplings to two 
photons and to $W$-bosons, confronting a precise prediction for this 
{\it observable} with 
results of experimental measurements should allow for stringent 
constraints on the deviations of these couplings from their Standard Model 
values. 

{\bf Acknowledgments}

We are grateful to Joey Huston for the clarification 
of some details of the ATLAS analysis.
We thank the Mainz Institute for Theoretical Physics (MITP) for
hospitality and partial support during the program \emph{Higher Orders and Jets for LHC}.

\end{document}